\documentclass[prl,twocolumn,secnumroman,tightenlines,amssymb,aps,nobibnotes,
               superscriptaddress,showpacs,balancelastpage,floatfix,nofootinbib]{revtex4}
\usepackage{graphicx}
\usepackage{multirow}           
\usepackage{color}	
\usepackage{ulem}
\usepackage[utf8]{inputenc}
\expandafter\ifx\csname package@font\endcsname\relax\else
\expandafter\expandafter
\expandafter\usepackage
\expandafter{\csname package@font\endcsname}
\fi
\DeclareRobustCommand\substyle{\name@idx{document substyle}}
\DeclareRobustCommand\classoption{\name@idx{document class option}}
\DeclareRobustCommand\classname{\name@idx{document class}}
\def\name@idx#1#2{{\ttfamily#2}
\index{#2\space#1=\string\ttt{#2}\space#1}\index{#1>#2=\string\ttt{#2}}}

\begin{document}

\title{Evidence for the general dominance of proton shells in low-energy fission}
\author{K. Mahata}
\email{kmahata@barc.gov.in}
\affiliation{Nuclear Physics Division, Bhabha Atomic Research Centre, Mumbai - 400085, India.}
\affiliation{Homi Bhabha National Institute, Anushaktinagar, Mumbai - 400094, India.}
\author{C. Schmitt}
\email{christelle.schmitt@iphc.cnrs.fr}
\affiliation{Institut Pluridisciplinaire Hubert Curien, 23 rue du Loess, B.P.\,28, 67037 Strasbourg Cedex 2, France}
\author{Shilpi Gupta}
\affiliation{Nuclear Physics Division, Bhabha Atomic Research Centre, Mumbai - 400085, India.}
\affiliation{Homi Bhabha National Institute, Anushaktinagar, Mumbai - 400094, India.}
\author{A. Shrivastava}
\affiliation{Nuclear Physics Division, Bhabha Atomic Research Centre, Mumbai - 400085, India.}
\affiliation{Homi Bhabha National Institute, Anushaktinagar, Mumbai - 400094, India.}
\author{G. Scamps}
\affiliation{Institut d'Astronomie et d'Astrophysique, Universite Libre de Bruxelles, Campus de la Plaine CP 226, 1050 Brussels, Belgium}
\author{K.-H. Schmidt}
\affiliation{Rheinstraße 4, 64390 Erzhausen, Germany}

\date{\today}

\begin{abstract}
\noindent
A regular pattern, revealing the leading role of the light-fragment nuclear charge, is found to emerge from a consistent analysis of the experimental information collected recently on low-energy asymmetric fission of neutron-deficient nuclei around lead. The observation is corroborated by a theoretical investigation within a microscopic framework, suggesting the importance of proton configurations driven by quadrupole-octupole correlations. This is in contrast to the earlier theoretical interpretations in terms of dominant neutron shells. The survey of a wider area of the nuclear chart by a semi-empirical approach points to the lack of understanding of the  competition between the different underlying macroscopic and microscopic forces in a quantitative manner. Combined with previously identified stabilizing forces, the present finding shows a striking connection between the ``old" (actinide) and ``new" (pre-actinide) islands of asymmetric fission which could steer the strive for an unified theory of fission.
\end{abstract}
\maketitle

{\it Introduction.} Fission is among the most dramatic examples of nuclear decay whereby a heavy nucleus splits into two fragments of comparable mass. Its discovery in the 1930's came as a surprize to the community, and recognition required the irrefutable chemical and physical evidence to be established \cite{hahn:1939, frisch:1939}. The fission process is important for various fields, including fundamental physics, astrophysics, and applied science \cite{andreyev:2018}.\\
Although fission was unexpected, an explanation for its high probability in heavy nuclei came quickly. It appeared that this re-arrangement of more than 200 nucleons - which is {\it a priori} a complex many-body quantum-mechanical problem, can be explained in an essentially classical way \cite{bohr:1939, meitner:1939}, in analogy with the division of a macroscopic liquid drop (LD) like a living cell. However, the LD model could not explain why, at low excitation energy, typical actinides predominantly split in an asymmetric manner \cite{flynn:1975, itkis:1990s}. The explanation had to await the late 1950's \cite{fong:1956}, and namely the introduction by Strutinsky of a method accounting for shell-structure effects in the calculation of the potential energy of the fissioning-system \cite{strutinsky:1966}, creating a complex landscape. It was then realized \cite{moller:1972, mosel:1992} that strong fragment-driven microscopic stabilization effects can supersede the gently evolving compound nucleus (CN) macroscopic energy, and dig deep valleys towards scission.\\
Natural candidates for producing fission valleys are the ``standard" magic numbers \cite{goeppert:1948}. For fission of actinides around uranium, fragment mass distributions of limited resolution (which constituted the main source of information for several decades) were interpreted as being due to the influence of shell effects, and namely neutron shells \cite{andreyev:2018}. The heavy-fragment mass distribution exhibiting a broad and robustly sitting structure around $A_H \approx$ 140 was found to be made up of two ``standard" contributions: the so-named $S1$ mode located at $A_H \approx$ 134, and the $S2$ mode at $A_H \approx$ 144, were ascribed to the neutron $N$ = 82 spherical, and a $N \approx$ 88 deformed, shell \cite{wilkins:1976}, respectively.\\
At the beginning of the century, a novel experimental method by Schmidt et al. \cite{schmidt:2000} revealed, however, that the $A_H \approx 140$ peak is characterized by a constancy of the heavy-fragment {\it charge} number at $Z_H \approx$ 52 and 55 for the $S1$ and $S2$ mode, respectively, with no preferential population of known neutron shells \cite{boeckstiegel:2008}. The $S1$ mode is characterized by a high $TKE$, a low neutron multiplicity for the heavy partner, and its yield increases with $N_{CN}/Z_{CN}$ approaching that of $^{132}$Sn \cite{andreyev:2018, schmidt:2016, waggemans:xxxx}. These observations  suggested that the $S1$ mode is primarily driven by the $Z$ = 50 shell aided by $N$ = 82. It is supported by the abrupt transition from asymmetric to symmetric fission while approaching $^{264}$Fm \cite{hulet:1986,wild:1990}. An interpretation for the $S2$ mode  was proposed recently only, in terms of the favored formation of a stabilized octupole configuration \cite{scamps:2018}.\\
Clear evidence of asymmetric fission in pre-actinides near $\beta$ stability, namely in $^{201}$Tl, was established in light-ion induced fission by Itkis et al. \cite{itkis:1990s}. The  recent observation of  almost exclusively asymmetric fission in the neutron-deficient $^{180}$Hg, for which two doubly-magic $^{90}$Zr fragments were expected \cite{andreyev:2010} has led to a resurgence of interest in fissioning pre-actinides, and measurements of additional systems \cite{ghys:2014, gorbinet:2014, nishio:2015, prasad:2015, igort:2018, gupta:2019, guptab:2019} ascertained the occurence of asymmetric fission over an enlarged domain around lead.\\
Finally, a consistent understanding of the fragment properties running from the ``old" (actinide) to the ``new" (pre-actinide) region of asymmetric fission is still missing \cite{andreyev:2018, schmidt:2019}. Advanced theories \cite{ichikawa:2018, scamps:2018, scamps:2019, schmidt:2019} have proposed different mechanisms to find an analogous origin. In this letter, the experimental information collected during the last few years in the neutron-deficient region around lead is analyzed in detail with the aim to elucidate its asymmetric fission properties, address the question of its origin, and seek a connection between the ``old" and the ``new" islands of asymmetric fission.\\ 
\\
{\it Method.}
The low excitation energy required for studying asymmetric fission of neutron-deficient pre-actinides can be ideally reached in $\beta$-delayed and electromagnetic-induced fission. Unfortunately, the number of systems accessible to these approaches is limited. A worldwide effort is therefore invested since a few years based on the alternative fusion-induced fission approach. The drawback of the method is the excitation energy imparted to the CN, which implies a weakening of possible shell-driven effects. The experimental information from the various approaches are analyzed in a common framework. The asymmetric components, clearly visible in the $\beta$-delayed \cite{andreyev:2010, ghys:2014} and electromagnetic-induced \cite{gorbinet:2014} fission experiments, in terms of the mean position of the light and heavy partners, in either mass $A_{L,H}$ or charge $Z_{L,H}$ depending on availability, are deduced from the measured fragment $A$ (or $Z$) distributions. As for the fusion-induced fission approach, the location of the asymmetric peaks was  determined in the corresponding references \cite{nishio:2015, prasad:2015, igort:2018, gupta:2019, guptab:2019} based on the adjustment of the mass distribution by a superposition of Gaussian functions. The mean neutron and/or proton numbers of the fission partners are then derived in this work under the Unchanged-Charge-Density (UCD) assumption \cite{vandebosch:1973}. The uncertainty (wherever applicable) introduced by this assumption was taken to be 0.8 unit.\\

\begin{figure}
\includegraphics[width=0.5\textwidth]{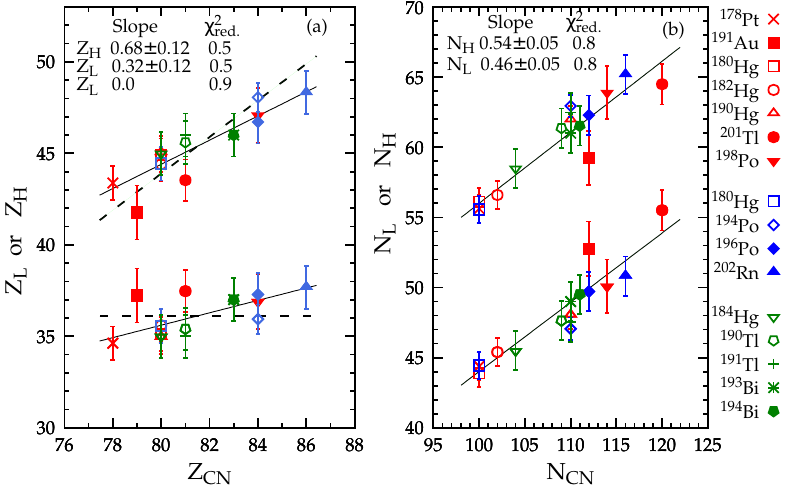}
\vspace{-.50cm}
\caption{Mean (a) proton $Z_{L(H)}$ and (b) neutron $N_{L(H})$ numbers of the light (heavy) fragment as a function of fissioning system $Z_{CN}$ and $N_{CN}$, respectively, from fusion- (red), $\beta$-delayed- (blue) and electromagnetic-induced (green) fission~\cite{itkis:1990s, andreyev:2010, ghys:2014, gorbinet:2014, nishio:2015, prasad:2015, igort:2018, gupta:2019, guptab:2019}. The best fits and $Z_L$= 36 are shown as continuous and dashed lines, respectively.} 
\label{fig1}
\end{figure}

\begin{figure}[!hbt]
\hspace{-1.cm}
\includegraphics[width=\linewidth]{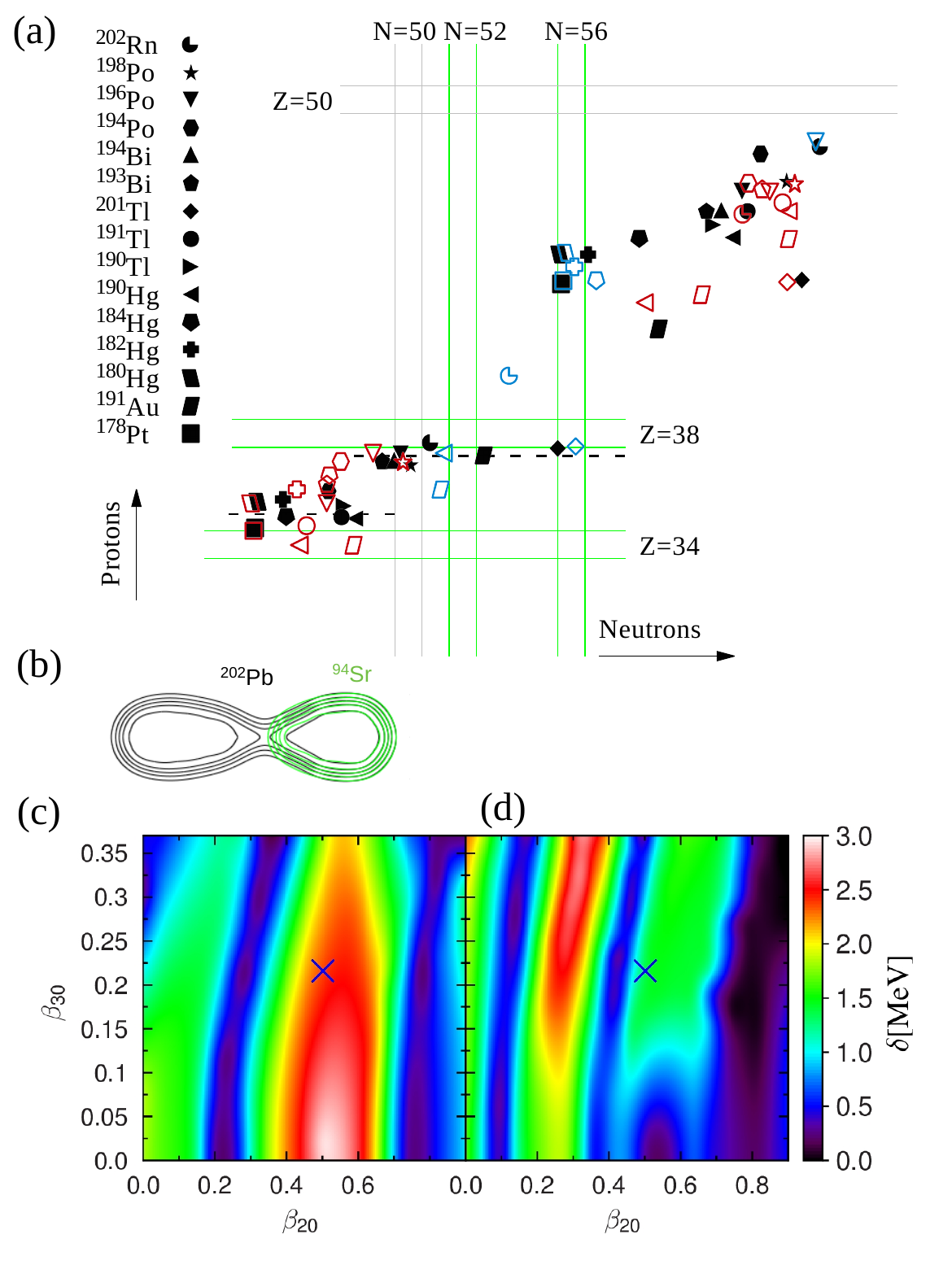}
\caption{(a) Comparison between the experimental  $Z_{L(H)}$ and $N_{L(H)}$ values (filled black symbols) and calculations (corresponding open symbols) with the EDF framework of Ref.~\cite{scamps:2019}. The blue and the red colors represent  compact and  elongated shapes of the fragments, respectively. The average experimental trends in Z$_L$ for N$_L$ below and above $\approx$50 are shown in black dashed lines. (b) Identification of the nascent light fragment with the quadrupole-octupole deformed $^{94}_{38}$Sr$_{56}$ density profile (green contour) in the energetically most favorable scission shape of $^{202}$Pb (black contour). Proton $Z$ = 38 (c) and neutron $N$ = 56 (d) shell gap $\delta$ in the ($\beta_{20}$, $\beta_{30}$) plane. Blue crosses correspond to the  scission shape of (b).}  
\label{fig2}
\end{figure}

{\it Results.}  To elucidate the nature of the asymmetric split in the pre-actinide region, the deduced $Z_{L,H}$ and $N_{L,H}$ are displayed in Fig.~\ref{fig1} as a function of, respectively, the total number of available protons ($Z_{CN}$) and neutrons ($N_{CN}$) in the fissioning system. Distinctly different behaviors, in the way the neutrons and protons are shared, are observed. Interestingly, the light-fragment charge $Z_{L}$ is seen to be confined within a narrow range around 36. Comparison between the $\chi^2$ of the free (full) and horizontally-constrained (dashed) adjustment of the $Z_L$ points shows that the slope is not very significant statistically. On the contrary, the heavy-fragment charge $Z_{H}$ exhibits a much stronger dependence on $Z_{CN}$, while both $N_{L,H}$ increase monotonically with increasing number of neutrons $N_{CN}$ to be shared. \\
It is evident from the rather stable location of $Z_L$, inferred from the present investigation involving a large diversity in ($A_{CN}$, $Z_{CN}$), that the light-fragment proton configuration plays the leading role in governing asymmetric fission of neutron-deficient nuclei around lead. This is at variance with previous interpretation which suggested the leading role played by neutrons (see {\it e.g.} Refs.~\cite{mulgin:1998, ichikawa:2018,scamps:2019,gorbinet:2014}). A preliminary survey \cite{schmidt:2019} based on four data points suggested that $Z_L$ indeed plays a specially intriguing role. Though no preference for specific neutron sharing can be observed, it is interesting to note in Fig.~\ref{fig1}(a) that the $Z_L$ values for  $N_L \gtrsim$ 50 are consistently higher ($\approx$ 37) than those ($\approx$ 35) for $N_L <$ 50 (see also  Fig.~\ref{fig2}). This is the primary reason for the observed small slope in $Z_L$ as a function of $Z_{CN}$.\\ 

{\it Comparison with theory.} State-of-the-art model calculations in the field provide a reasonable description of asymmetric fission of $^{180}$Hg \cite{panebianco:2012, warda:2012, ichikawa:2012, scamps:2019}. Ichikawa and Moller \cite{ichikawa:2018}, based on a macroscopic-microscopic approach, attribute the observed mass asymmetry to a  shell gap developing at the outer saddle point in the neutron sub-system. Using the microscopic energy density functional (EDF) framework, Scamps and Simenel~\cite{scamps:2019} have concluded the dominance of octupole effects, in most cases driven by neutron configurations. Another interpretation  \cite{warda:2012} relate the asymmetric splits to prescission configurations involving molecular structures, and namely a spherical $^{90}$Zr.\\
For the systems presented in this letter we have performed calculations within the quantum-based EDF theory of Ref.~\cite{scamps:2019}. The experimental  $Z_{L (H)}$ and $N_{L (H)}$ are compared in Fig.~\ref{fig2} with the predicted corresponding most probable values as determined by the bottom of the asymmetric fission valley just before scission. For odd-nuclei, the calculation corresponds to the average from the neighboring even-even systems. A reasonably good agreement between the experimental data and theory is observed. In particular, the relative constancy of $Z_L$ is reproduced, highlighting the dominance of proton configurations over the neutron effects proposed in Ref.~\cite{scamps:2019}. Interestingly, the calculation seems also to exhibit a bunching of $Z_L$ into two subgroups, depending on whether $N_L$ is below or above $\approx 50$. In Ref.~\cite{scamps:2019}, a detailed analysis of neutron-deficient system (namely $^{178}$Pt) identified a shell gap at $Z_L$ = 34 for $N_L <$ 50 caused by a stabilized large quadrupole-octupole deformation at scission from the single particle energy levels of $^{78}_{34}$Se$_{44}$. Intrigued by the present observation, a similar detailed analysis was performed for a less neutron-deficient system {\it i.e.},$^{202}$Pb. Another shell gap was seen to be present at $Z_L$ = 38 when $N_L >$ 50 and corresponding to a comparatively more compact deformation. The outcome of this new analysis is illustrated in Fig.~\ref{fig2}. From the identification technique proposed in Ref.~\cite{warda:2012} improved by taking into account the octupole degree of freedom \cite{scamps:2018}, we compare in panel (b) the density profile of the quadrupole-octupole deformed $^{94}_{38}$Sr$_{56}$ with the nascent light fragment in fission of $^{202}$Pb.  The last row of the figure displays the magnitude of the (c) proton and (d) neutron shell gap $\delta$ for this fragment in the ($\beta_{20}$, $\beta_{30}$) deformation plane. Interestingly, even though a large shell gap for $N$ = 56 is present at smaller deformations ($\beta_{20}\sim$0.3), the proton $Z$ = 38 gap is found to dominate for the identified scission shape. This corroborates that stabilized deformed proton configurations  play the dominant role  in deciding the fission partition in pre-actinides, where theory was originally \cite{scamps:2019} anticipating neutrons to play the leading role.\\


{\color{blue}

}


\begin{figure*}[!hbt]
\includegraphics[trim=0 0 0 0,clip,width=\textwidth,angle=0]{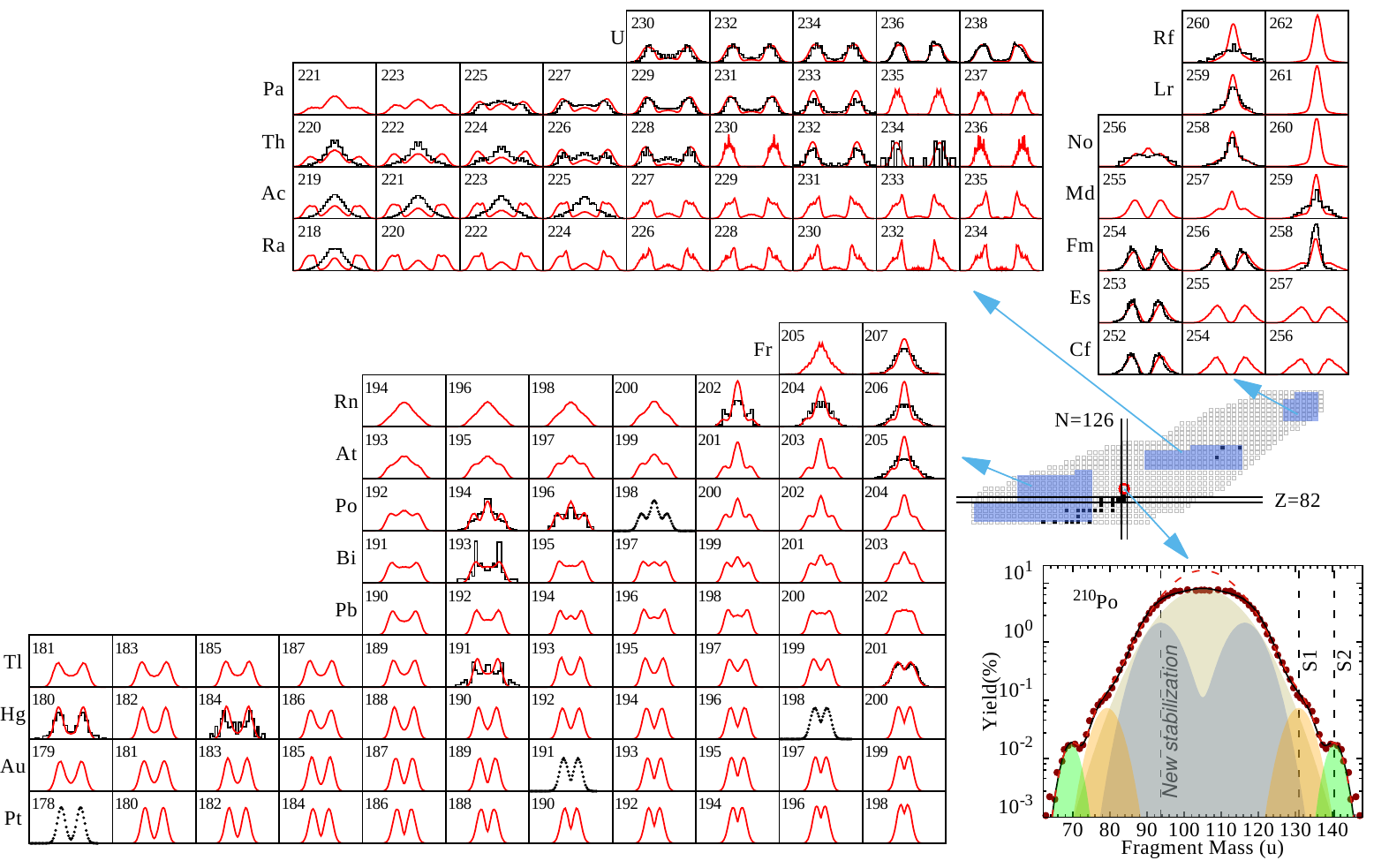}
\caption{ Experimental fragment mass distributions across the nuclear chart~\cite{itkis:1990s, andreyev:2010, ghys:2014, nishio:2015, prasad:2015, igort:2018, gupta:2019, guptab:2019,schmidt:2000, gorbinet:2014,nishio:mnt,refSF} at low energy fission (black histograms).  Wherever the fragment $Z$ was measured, mass was obtained under the UCD assumption. The calculations by the code GEF2019/V1.1 (red full lines) are done at the actual $E^*$ for the measured systems.   For fusion-fission, the GEF result obtained at low energy is shown (black full dots), extrapolated from the overall reasonable description by the model of measurements done at intermediate energy \cite{guptab:2019}. The inset shows the multi-Gaussian fit, using the ``old" and ``new" fission modes, to the experimental mass distribution of $^{210}$Po at an excitation energy 9.5 MeV above the fission barrier~\cite{itkis:1990s}, see the text. The red dashed line represent the fit without the ``new" mode.
} 
\label{fig4}
\end{figure*}

{\it Fission-fragment properties over the nuclear chart.} The pronounced double-humped mass distribution of $^{180}$Hg being reminiscent of the distribution measured around uranium, it is very intriguing to search for a connection between the ``old" and ``new" isolated regions of asymmetric fission.\\
The evolution of the fragment mass distribution as measured across the nuclear chart for low-energy fission of isotopes between platinium ($Z_{CN}$=78) and rutherfordium ($Z_{CN}$=104) is illustrated in Fig.~\ref{fig4}. To cover an as wide as possible domain, the results from various experimental methods are included. That implies some spread ($\sim$10MeV) in the initial excitation energy of the fissioning system. However, this spread does not impact the present discussion as the overall shape of the distributions does not change significantly over such a energy range \cite{schmitt:2018}.\\ 
Experimentally, an asymmetric fission component starts to be visible in the actinide region for $A_{CN}$ above $\approx$ 224, and persists up to $A_{CN} \approx$ 256. On either side of this domain, symmetric fission dominates. On the right side, the dominance of symmetric fission is driven by the summing-up of shell effects in the population of two fragments around $^{132}$Sn in the heaviest transfermiums.  At the left of the domain (from radon to radium), symmetric fission prevails for the lightest actinides due to the dominant influence of the macroscopic potential which outweights the $Z_H$ quantum effect that governs $S1$ and $S2$~\cite{benlliure:1998}. Towards the neutron-deficient region around lead, southwest of the traditional actinides, asymmetric fragmentation abounds again. As established in this work, there the leading effect is  played by $Z_L \approx$ 36.\\
\\
In the absence of calculations by a fundamental theory over the wide domain of Fig.~\ref{fig4}, we consider the semi-empirical GEneral Fission model (GEF2019/V1.1) \cite{schmidt:2016, schmidt:2019}. The GEF formalism casts the essential ideas about the physics of fission, including the new stabilization in the preactinide region, into simplified equations with a global set of parameters, obtained by fitting benchmark experimental data. The calculated mass distributions are shown in Fig.~\ref{fig4} with red full lines. Overall, the model performs impressively well; in particular, the most neutron-deficient pre-actinides, heavy actinides and the fermium region are nicely described. Thus, its predictions in the pre-actinide region, where the experimental information is still limited, can be considered as of reliable guidance for future studies. However, in the region of most neutron-deficient radium to thorium, GEF fails to reproduce the experimental observations, indicating that the competition between the structural effect(s) at play in a specific region and the macroscopic force is not fully understood yet. The detailed theoretical models~\cite{mumpower:2020,zhao:2019,regnier:2019} also fail to reproduce the experimental observations in this region.\\

Since quantum effects are a property of the nucleus {\it per se}, if the nuclear structure of the nascent fragments indeed plays a key role, analogous stabilizations {\it must} be at play in the pre- and actinide islands. The manifestation of a specific configuration in fission yields of a given system is then a matter of competition. A weak persistence of the $S1$ and $S2$ modes in pre-actinides between $^{205}$Bi and $^{213}$At was indeed seen by Itkis et al.~\cite{itkis:1990s}, progressively dying out with increasing neutron-deficiency. The observation of sizeable asymmetric components more close to symmetric split~\cite{itkis:1990s}  was earlier attributed to neutron shells \cite{mulgin:1998}. According to the outcome of the present investigation, this finding is instead proposed to be governed by the influence of the here-evidenced $Z_L$ stabilization, and which is aided by specific $N$ configurations with increasing $N_{CN}$ \cite{mulgin:1998, wilkins:1976}. The inset of fig.~\ref{fig4} shows that the experimental data on $^{210}$Po \cite{itkis:1990s}, which is ideally situated at the crossroads of the old and new islands, can be accurately described once all the identified effects, which we attribute to proton configurations, are taken into account.\\

\begin{figure}[!hbt]
\hspace{-1.cm}
\includegraphics[width=9.2cm, height=5.8cm,angle=0]{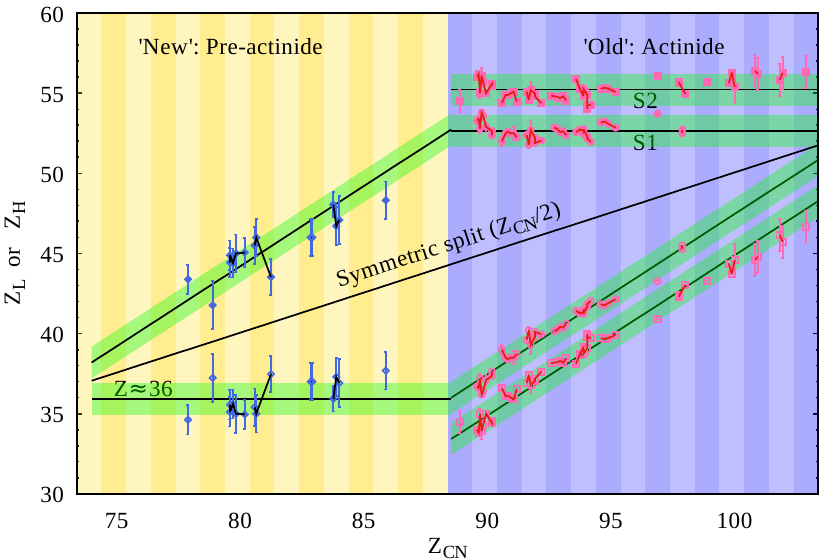}
\caption{Evolution of the average $Z_{L(H)}$ positions for the asymmetric fission channel as a function of $Z_{CN}$ from above rare-earth to very heavy elements. For clarity, isotopes of a same element are shifted according to mass. The points are from Refs.~\cite{boeckstiegel:2008,hulet:1986,wild:1990,itkis:1990s, andreyev:2010, ghys:2014, gorbinet:2014, nishio:2015, prasad:2015, igort:2018, gupta:2019, guptab:2019} and therein.}
\label{fig6}
\end{figure}

Finally, the evolution of the $Z_{L (H)}$ peak positions from above rare-earth to very heavy elements is presented in Fig.~\ref{fig6}, pointing to a natural connection between the two islands of asymmetry. As discussed earlier, in the pre-actinide region the light fragment position remains fixed at $Z \approx$ 36, and the heavy fragment charge increases steadily till it merges with the $S1$ mode at actinium. Thereafter, $S1$ and $S2$ start to dominate, forcing the light fragment to increase its charge with $Z_{CN}$. It is interesting to note that the role reversal of the light and heavy fragment occurs exactly at the boundary between pre-actinides and actinides. The observed geometric connection between these two islands establishes the general dominance of proton shells in low-energy fission. It raises fundamental  questions like ``Why is the influence of the protons so dominant in the sharing of nucleons in fission?", ``Is the neutron subsystem strongly disturbed by the neck, while the protons are pushed away into the fragments by the Coulomb repulsion allowing specific stabilized configurations to manifest?". The present finding emphasizes the need for further experimental studies, both near the reversal boundary and towards the limits. On the low $Z_{CN}$ side, it would in particular be interesting to see if fission becomes symmetric in the region around neutron-deficient hafnium ($Z_{CN}$ = 2 $\times$ 36), as predicted in Ref.~\cite{scamps:2019}. On the upper $Z_{CN}$ side, the abrupt resurgence of predominantly symmetric fission was already seen in heavy fermium-like elements. For still larger $Z_{CN}$, linear continuity  would suggest another role reversal, the light fragment formation being this time driven by the standard ($S1$ and $S2$) fission modes. Such a picture was recently predicted within a macroscopic-microscopic prescission point model \cite{carjan:2019}. Dynamical calculations based on either a microscopic mean-field  \cite{matheson:2019} or a macro-microscopic \cite{albertsson:2020} approach instead predict a more asymmetric partitioning related to $^{208}$Pb cluster radioactivity, and the matter is vividly debated \cite{usang:2020}. It is interesting to note that, for the super-heavy nuclei, the lead fragment cluster would be associated with a light partner having $Z_L \approx$ 36.\\
\\
\\
{\it Conclusion.} The consistent analysis of the experimental information  on fission of neutron-deficient nuclei around lead reveals the leading role played by the light fragment proton configuration, which is in contrast to the predicted dominance of neutron shells in previous theoretical studies. Detailed theoretical investigation using the microscopic EDF framework attribute the experimental observation to shell stabilizations at Z = 34 and 38 associated with more elongated shape ($\beta_2\gtrsim$ 0.50) for  $N_L <$ 50 and more compact shape ($\beta_2\lesssim$ 0.50) for $N_L \gtrsim$ 50, respectively. Combining a light-fragment-driven stabilizing effect identified in this work with previously established leading effects, a striking connection between the old and new islands of asymmetric fission is found to hold, explaining the main trends from above rare-earth to very heavy element fission in an essentially ``simple" way for the first time, and  establishing the general dominance of proton shells in low-energy fission. 
Large-scale calculations within a semi-empirical model suggest that a major deficiency of current understanding is in the complex and quantitative interference of the various macroscopic and microscopic forces. 
The experimental evidence made available in this work is essential for addressing this question, and guiding the necessary development towards a unified theory of fission. \\  
\\
\\
{\bf Acknowledgements}
\\
One of the authors (CS) thanks Andrei Andreyev, Peter Möller and Cedric Simenel for enlighting discussions. CS acknowledges financial support under the French-Indian LIA, and the GSI/IN2P3-CNRS collaboration 19-80.\\
\\
\\


\begin{thebibliography}{99}
\bibitem{hahn:1939} O. Hahn and F. Strassmann, Naturwissenschaften 27 (1939) 11.
\bibitem{frisch:1939} O. Frisch, Nature 143 (1939) 276.
\bibitem{andreyev:2018} A. N. Andreyev, K. Nishio, and K.-H. Schmidt, Rep. Prog. Phys. 81 (2018) 016301.
\bibitem{bohr:1939} N. Bohr and J.A. Wheeler, Phys. Rev. 56 (1939) 426.
\bibitem{meitner:1939} L. Meitner and O. Frisch, Nature 143 (1939) 239.
\bibitem{flynn:1975} K.F. Flynn et al., Phys. Rev. C 5 (1972) 1725.
\bibitem{itkis:1990s} M.G. Itkis et al., Sov. J. Nucl. Phys. 41 (1985) 544, 43 (1986) 1125, 47 (1988) 4, 52 (1990) 601 and 53 (1991) 757.
\bibitem{fong:1956} P. Fong, Phys. Rev. 102 (1956) 434.
\bibitem{strutinsky:1966} V.M. Strutinsky, Nucl. Phys. A 95 (1967) 420.
\bibitem{moller:1972} P. Moller,  Nucl. Phys. A 192 (1972) 529.
\bibitem{mosel:1992} U. Mosel and H.W. Schmitt, Nucl. Phys. A 165 (1971) 73 and Phys. Rev. C 4 (1971) 2185.
\bibitem{goeppert:1948} M. Goeppert-Mayer, Phys. Rev. 74 (1948) 235.
\bibitem{wilkins:1976} B.D. Wilkins, E.P. Steinberg, and Chasman, Phys. Rev. C 14 (1976) 1832. 
\bibitem{schmidt:2000} K.-H. Schmidt et al., Nucl. Phys. A 665 (2000) 221.
\bibitem{boeckstiegel:2008} C. Boeckstiegel et al., Nucl. Phys. A 802 (2008) 12.
\bibitem{waggemans:xxxx} L. Dematte et al. Nucl. Phys. A 617 (1997) 331.
\bibitem{schmidt:2016} K.-H. Schmidt et al., Nucl . Data Sheets 131 (2016) 107.
\bibitem{hulet:1986} E.K. Hulet et al., Phys. Rev. Lett.  56 (1986) 313.
\bibitem{wild:1990} J.F. Wild et al., Phys. Rev. C 41 (1990) 640.
\bibitem{scamps:2018} G. Scamps and C. Simenel, Nature (London) 564 (2018) 382.
\bibitem{andreyev:2010} A. Andreyev et al., Phys. Rev. Lett 105 (2010) 252502.
\bibitem{ghys:2014} L. Ghys et al., Phys. Rev. C 90 (2014) 041301.
\bibitem{gorbinet:2014} T. Gorbinet et al., Scientific Workshop on Nuclear Fission dynamics and the Emission of Prompt Neutrons and and $\gamma$-rays, Opatcja, Croatia (Sept 2014).
\bibitem{nishio:2015} K. Nishio et al., Phys. Lett. B 748 (2015) 89.
\bibitem{prasad:2015} E. Prasad et al., Phys. Rev. C 91 (2015) 064605.
\bibitem{igort:2018} I. Tsekhanovich, Phys. Lett. B 790 (2019) 583.
\bibitem{gupta:2019} S. Gupta et al., Phys. Lett. B 803 (2020) 135297.
\bibitem{guptab:2019} S. Gupta et al., Phys. Rev. C 100 (2019) 064608.
\bibitem{schmidt:2019} K.-H. Schmidt and B. Jurado, Rep. Progr. Phys. 81 (2018) 106301.
\bibitem{ichikawa:2018} T. Ichikawa and P. Moller, Phys. Lett. B 789 (2019) 679.
\bibitem{scamps:2019} G. Scamps and C. Simenel, Phys. Rev. C 100 (2019) 041602, and Supplemental Material.
\bibitem{vandebosch:1973} R. Vandebosch and J.R. Huizenga, {\it Nuclear fission} (Academic, New York, 1973).
\bibitem{mulgin:1998} S.I. Mulgin et al., Nucl. Phys. A 640 (1999) 375. 
\bibitem{benlliure:1998} J. Benlliure et al., Nucl. Phys. A 628 (1998) 458. 
\bibitem{panebianco:2012} S. Panebianco et al., Phys. Rev. C 86 (2012) 064601.
\bibitem{warda:2012} M. Warda et al., Phys. Rev. C 86 (2012) 024601.
\bibitem{ichikawa:2012} T. Ichikawa et al., Phys. Rev. C 86 (2012) 024610.
\bibitem{schmitt:2018} C. Schmitt et al., EPJ Web of Conferences 169, 00022 (2018).
\bibitem{mumpower:2020} M. R. Mumpower et al., Phys. Rev. C 101 (2020) 054607.
\bibitem{zhao:2019} J. Zhao et al., Phys. Rev. C 99 (2019) 054613.
\bibitem{regnier:2019} D. Regnier et al., Phys. Rev. C 99 (2019) 024611.
\bibitem{nishio:mnt} K. Hirose et al., Phys. Rev. Lett. 119 (2017) 222501.
\bibitem{refSF} E. K. Hulet et al., Phys.Rev. C 40 (1989) 770 and therein.
\bibitem{carjan:2019} N. Carjan et al., Phys. Rev. C 99 (2019) 064606.
\bibitem{matheson:2019} Z. Matheson et al., Phys. Rev. C 99 (2019) 041304(R).
\bibitem{albertsson:2020} M. Albertsson et al., Eur. Phys. J. A 56 (2020) 46. 
\bibitem{usang:2020} C. Ishizuka et al., Phys. Rev. C 101 (2020) 011601(R). 

\end{thebibliography}
\end{document}